\documentclass{IEEEtran}
\pdfoutput=1
\usepackage[T1]{fontenc}
\usepackage[utf8]{inputenc}

\usepackage{ifthen}

\usepackage{amsmath}
\usepackage{amssymb}
\usepackage{comment}
\usepackage{booktabs}
\usepackage{graphicx}
\usepackage[nolist]{acronym}

\usepackage[caption=false,font=footnotesize]{subfig}

\usepackage{tikz}
\usepackage{pgfplots}

\usetikzlibrary{arrows,calc,positioning}

\usepgfplotslibrary{groupplots}

\makeatletter
\@ifundefined{ifstandalone}{}{
\ifstandalone
  \setlength{\linewidth}{9cm}
\fi
}
\makeatother

\def\oh{\tfrac{1}{2}}

\newcommand{\mat}[1]{\mathbf{#1}}
\newcommand{\C}{^{\text{C}}}

\def\changesdir{changes/}
\newif\ifbypass
\bypasstrue  
\usepackage{environ}

\ifx\changesdir\undefined
\PackageError{review}{Directory for changes not defined. Define the
  variable ``changesdir'' before including this package. Make sure to
  end with a forward slash. Also make sure, the directory exists}{}
\fi

\newwrite\reviewcomment
\makeatletter
\NewEnviron{newchangeE}[1][]{%
\ifbypass%
\BODY%
\else
\ifx&#1&%
\BODY%
\else%
\typeout{Writing the paragraph also to the file #1}%
\def\fnname{\changesdir#1.comment}
\immediate\openout\reviewcomment=\fnname%
\immediate\write\reviewcomment%
{\noexpand\setcounter{equation}{\expandafter\arabic{equation}}}%
\toks@=\expandafter{\BODY}%
\immediate\write\reviewcomment{\the\toks@}%
\immediate\closeout\reviewcomment%
\input{\fnname}%
\fi%
\fi}
\makeatother

\newcommand{\newchange}[2][]{\begin{newchangeE}[#1]#2\end{newchangeE}}

\ifbypass

\newcommand{\MMa}[2][]{#2}

\newcommand{\CR}[2][]{#2}
\else

\newcommand{\MMa}[2][]{\textcolor{brown}{[\newchange[#1]{#2}]}}

\newcommand{\CR}[2][]{\textcolor{magenta}{[\newchange[#1]{#2}]}}
\fi

\makeatletter
\def\maketag@@@#1{\hbox{\m@th\normalfont\normalsize#1}}
\makeatother

\usetikzlibrary{external}
\tikzstyle{ReIm}=[draw,minimum height=5mm,minimum width=7mm,inner sep=2pt]
\tikzstyle{Mult}=[draw,circle,inner sep=0]

\tikzstyle{shorten<>}=[shorten >=#1, shorten <=#1]

\definecolor{mygreen}{rgb}{0, 0.5, 0}

\gdef\R{\tikz \draw (0,0) circle (4.5pt);}
\gdef\I{\tikz \filldraw (0,0) circle (4pt); }

\usepackage{standalone}

\title{Conjugate-Root Offset-QAM for Orthogonal Multicarrier Transmission}

\author{
	\IEEEauthorblockN{Maximilian Matth\'{e}, Gerhard Fettweis}\\
}

\begin{document}
\begin{acronym}
  \acro{TFL}{time-frequency localization}
  \acro{BLT}{Balian-Low theorem}
  \acro{PPN}{poly-phase network}
  \acro{RRC}{root raised cosine}
  \acro{RC}{raised cosine}
  \acro{CR}{conjugate-root}
  \acro{CRRC}{conjugate RRC}
  \acro{CP}{cyclic prefix}
  \acro{ZF}{zero-forcing}
  \acro{MF}{matched filter}
  \acro{OOB}{out-of-band}
  \acro{SER}{symbol error rate}
  \acro{GFDM}{Generalized frequency division multiplexing}
  \acro{DFT}{discrete Fourier transform}
  \acro{FFT}{fast Fourier transform}
  \acro{ICI}{inter-carrier interference}
  \acro{PSD}{power spectral density}
  \acro{GS}{guard symbol}
  \acro{MC}{multicarrier}
\end{acronym}

\maketitle

\begin{abstract}
  Current implementations of OFDM/OQAM are restricted to band-limited
  symmetric filters.
  To circumvent this, non-symmetric
  conjugate root (CR) filters are proposed for OQAM modulation.  The system is applied
 to Generalized Frequency Division Multiplexing (GFDM) and a method
 for achieving transmit diversity with OQAM modulation is presented.
 The proposal reduces implementation complexity compared to existing
 works and provides a more regular phase space.
 GFDM/CR-OQAM outperforms conventional GFDM in terms of symbol
 error rate in fading multipath channels and provides a more localized
 spectrum compared to conventional OQAM.
\end{abstract}
\acused{GFDM}


\section{Introduction}

With%
\let\thefootnote\relax\footnote{Manuscript received..., revised\\
  The responsible editor during processing at IEEE WCL was ...\\
  Here goes some more editorial text...\\
  The DOI of this document is ....\\
  This work has been performed in the framework of the FP7 project
  ICT-619555 RESCUE, which is partly funded by the European Union.\\
  M. Matth\'e and G. Fettweis are with Vodafone Chair Mobile
  Communication Systems, Technische Universität Dresden, Germany
  (e-mail: maximilian.matthe@ifn.et.tu-dresden.de;
  fettweis@ifn.et.tu-dresden.de)} 5G on the horizon, new waveforms for
the PHY layer are investigated that are suitable for the upcoming
requirements \cite{Wunder2014}.  In particular, good \ac{TFL} of the
transmit signal is required to cope with asynchronicities
\cite{Floch95} and to provide a low \ac{OOB} radiation, which is
needed for spectral agility and carrier aggregation.  Filtered
\ac{MC} systems \cite{Chang1966} provide the means for good \ac{TFL}
by adaptation of the prototype filter and spectral agility is achieved
by switching on and off certain subcarriers.  High spectral efficiency
is important to serve the increased demand for high speed data access
\cite{Wunder2014}.  Hence, a future waveform should transmit symbols
at the Nyquist rate in order to not waste valuable time-frequency
resources.

Even in a distortion-free channel, the \ac{BLT}
prohibits the distortion-free reconstruction of
complex valued-symbols sent at Nyquist rate when using filters with
good \ac{TFL} \cite{Benedetto1998}.
Hence, in terms of \ac{SER}, QAM \ac{MC} systems with good \ac{TFL} perform
worse than orthogonal systems without good \ac{TFL}.
One possibility to circumvent the \ac{BLT} is to transmit real-valued
symbols at twice the Nyquist rate, which was initially proposed
in~\cite{Chang1966}.  This gave rise to the well known
OFDM/OQAM~\cite{Siohan2002} \ac{MC} system where both
orthogonality and good \ac{TFL} is kept by using Offset-QAM (OQAM),
which is proposed for upcoming 5G networks \cite{Wunder2014}.

\MMa[cSym1]{In current OQAM proposals, the employed real-valued
  prototype filter is required to be bandlimited and symmetric and a
  phase shift of $\tfrac{\pi}{2}$ is maintained between adjacent
  subcarriers. In \cite{Vangelista2001}, low-complexity
  implementations of these schemes are described and also a method is
  proposed to relax the requirement of symmetry of the employed
  prototype filter which is advantageous since it is shown in
  \cite{Feher1993}, that non-symmetric filters can be more robust
  against multipath fading. However, the approach presented in
  \cite{Vangelista2001} suffers from an increased implementation
  complexity compared to conventional OQAM since it requires to use
  two \acp{PPN}. }

In this paper we contribute another version of \ac{MC} OQAM
modulation, named \ac{CR} Offset-QAM (CR-OQAM), which allows to use
the class of non-symmetric CR filters, removing the $\tfrac{\pi}{2}$
phase shift and providing means for an easier implementation compared
to \cite{Vangelista2001}.
We further apply CR-OQAM to \ac{GFDM} \cite{Michailow2014}, which is another 5G candidate
waveform that normally uses QAM modulation and hence suffers from the
\ac{BLT}. With CR-OQAM, we provide a well-localized orthogonal system,
improving the \ac{SER} performance in fading multipath channels
compared to standard \ac{GFDM}.  Additionally, time-reversal
space-time-coding (TR-STC) \cite{Matthe2015a} is applied to
GFDM/CR-OQAM to exploit multi-antenna diversity, showing that in GFDM
STC can be combined with OQAM modulation which is a known issue
in OFDM/OQAM \cite{Lele2010}.  \MMa[Siohan]{In \cite{Lin2014},
  FBMC/COQAM is proposed, being equivalent to GFDM using
  OQAM. However, compared to the present paper, neither CR-OQAM nor
  STC is covered therein.}

The remainder is organized as follows: Sec. II describes the
conventional OQAM transceiver, whereas the proposed modification is
presented in Sec. III. Sec. IV describes the application of CR-OQAM to
GFDM and performance simulations are provided in Sec. V. Finally,
Sec. VI concludes.

\section{Conventional OFDM/OQAM}



\def\oqamheight{4cm}
\begin{figure*}[t]
  \centering
  \resizebox{!}{3.6cm}{\includegraphics{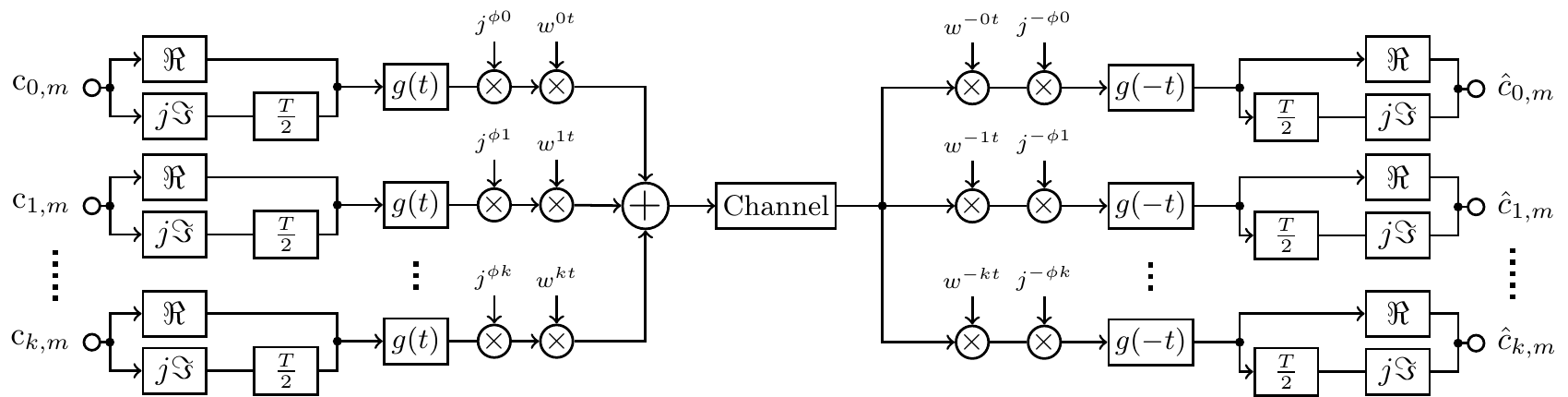}}
  \caption{Block diagram of the OFDM/OQAM ($\phi=1$) and OFDM/CR-OQAM
    ($\phi=0$) transceiver.}
  \label{fig:convOqam}
\end{figure*}

In OFDM/OQAM, complex-valued data symbols $c_{k,m}$ are transmitted on
K subcarriers, where real and imaginary part are offset by
$\tfrac{T}{2}$ where $T$ is the symbol duration.  Each symbol is
pulse-shaped with a symmetric, real-valued pulse shaping filter $g(t)$
\cite{Chang1966,Siohan2002}. The transmission equation is given by
\parbox[t]{\linewidth}{\small
\begin{align}
  x(t) &= \sum_{k=0\atop m\in\mathbb{Z}}^{K-1}(c_{k,m}^{\text{R}}g(t-mT)+jc_{k,m}^{\text{I}}g(t-mT-\tfrac{T}{2}))j^kw^{kt}
\end{align}
}
where $w=\exp(j2\pi F)$, $F=\tfrac{1}{T}$, and $c^{\text{R}}_{k,m}$ and
$c^{\text{I}}_{k,m}$ are real and imaginary part of $c_{k,m}$,
respectively. Note that, due to the factor $j^k$, adjacent subcarriers
differ by a phase shift of $\tfrac{\pi}{2}$.

At the receiver, matched filtering is carried out, i.e.
\begin{align}
  \hat{c}^{\text{R}}_{k,m} &= \Re (x(t)j^{-k}w^{-kt}*g(-t))|_{t=mT}\\
  \hat{c}^{\text{I}}_{k,m} &= \Im (x(t)j^{-k}w^{-kt}*g(-t))|_{t=(\oh+m)T},
\end{align}
where convolution $x(t)*y(t)$ is defined as
\begin{align}
  (x(t)*y(t))|_{t=\tau}&=\int_{-\infty}^{\infty}x(t)y(\tau-t)dt.
\end{align}
The corresponding block diagram of the OFDM/OQAM transceiver is
presented in Fig. \ref{fig:convOqam}. For perfect reconstruction, the
orthogonality conditions between the $k$th and $(k+\kappa)$th subcarrier
that are given by \cite[eq. (13)]{Siohan2002}
\begin{align}
\label{eq:o1} \Re& \bigl\{(g(t)j^{-\kappa}w^{-\kappa t}*g(-t))|_{t=mT\vphantom{\oh}}\bigr\}&&=\delta(\kappa,m)\\
\label{eq:o2} \Re& \bigl\{(jg(t)j^{-\kappa}w^{-\kappa t}*g(-t))|_{t=(\oh+m)T}\bigr\}&&=0\\
\label{eq:o3} \Im& \bigl\{(g(t)j^{-\kappa}w^{-\kappa t}*g(-t))|_{t=(\oh+m)T}\bigr\}&&=0\\
\label{eq:o4} \Im& \bigl\{(jg(t)j^{-\kappa }w^{-\kappa
    t}*g(-t))|_{t=mT\vphantom{\oh}}\bigr\}&&=\delta(\kappa ,m)
\end{align}
must be fulfilled for all $\kappa,m$.  With the relations
\mbox{$\Re\{ja\}=-\Im\{a\}$} and $\Im\{ja\}=\Re\{a\}$, (\ref{eq:o1})
and (\ref{eq:o2}) are equivalent to (\ref{eq:o4}) and (\ref{eq:o3}),
respectively.  These conditions are in particular fulfilled, when
$g(t)$ is a symmetric, half-Nyquist filter with band-limitation $G(f)=0,
|f|\geq{}F$ \cite{Floch95}.

\begin{figure*}[t]
  \centering
  \resizebox{!}{4.3cm}
  {
   \subfloat[\ac{ICI} of \ac{RRC}.]{\label{fig:adj}
   \includegraphics[width=0.33\linewidth]{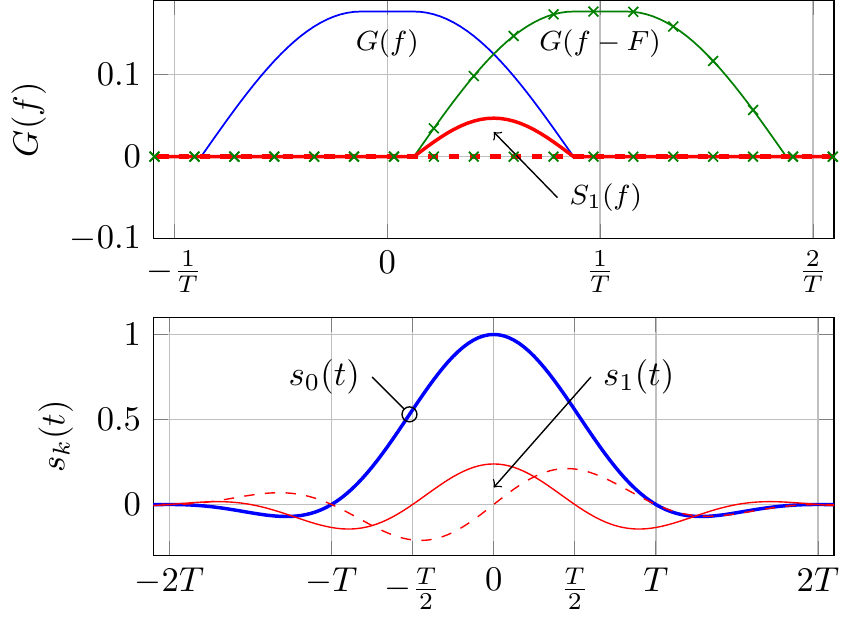}}
   \subfloat[\ac{RRC} and \ac{CRRC} impulse and frequency response.]{\label{fig:tdfd}
   \includegraphics[width=0.33\linewidth]{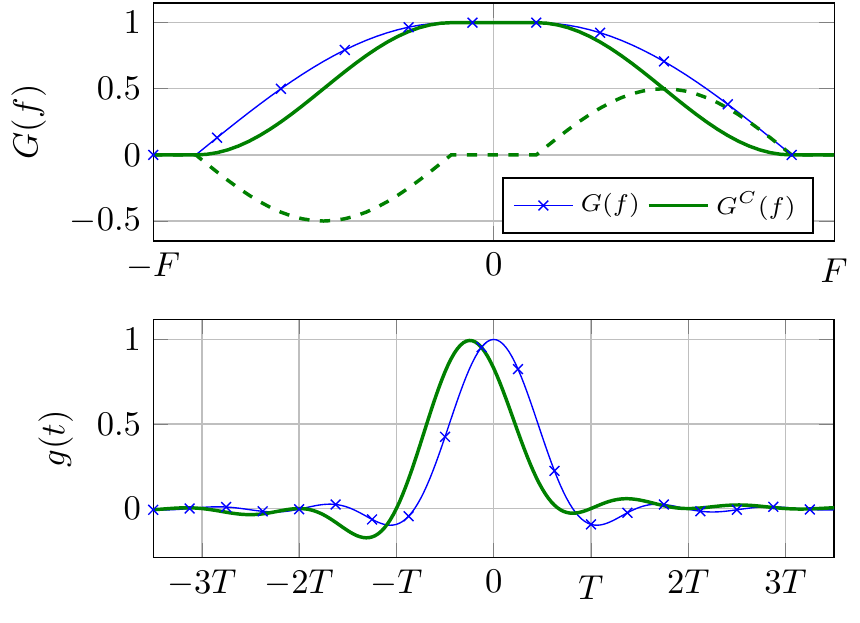}}
   \subfloat[\ac{ICI} of \ac{CRRC}.]{\label{fig:adjcr}
   \includegraphics[width=0.33\linewidth]{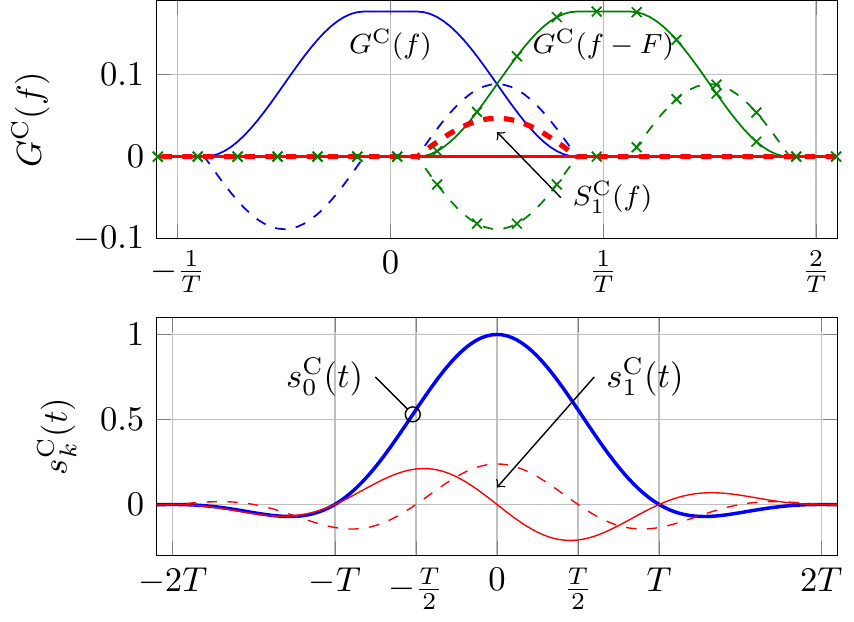}}}
 \caption{Solid and dashed lines represent real and imaginary part,
   respectively. (a), (c) ICI of adjacent channel. (b) Filter response of RRC and CRRC
   filters.}
  \label{fig:interference}
\end{figure*}

\noindent The convolution is expressed with the Fourier transform
$\mathcal{F}$ by
\begin{align}
  (g(t)w^{-kt}*g(-t))|_{t=\tau}&=\mathcal{F}^{-1}\{S_k(f)\}(\tau)=s_k(\tau),
\end{align}
where $S_k(f)=G(f-kF)G^*(f)$ is the spectrum of the intercarrier
interference (ICI) from the $k'$th to the $(k'+k)$th subcarrier,
$s_k(t)$ is the corresponding time domain function, $G(f)$ is the
Fourier transform of $g(t)$ and $(\cdot)^*$ denotes complex
conjugation.  Fig. \ref{fig:adj} depicts $s_1(t)$ for a
\ac{RRC} filter with roll-off $\alpha=0.75$.




\section{Conjugate-Root Multicarrier OQAM}
In literature, mostly symmetric half-Nyquist filters are employed for
\ac{MC} OQAM systems. We propose the application of
non-symmetric \ac{CR} filters \cite{Demeechai1998,Tan2004}
for \ac{MC} OQAM systems.



Assume an even $H(f)$ with band limitation $H(f)=0$ for
$|f|\geq F$. Then, $H(f)$ fulfills the 1st Nyquist criterion
\cite{Demeechai1998} iff
\begin{align}
  \label{eq:NQ}
  \forall  f\in[0,F]: H(f)+H(f-F)=1.
\end{align}
Its corresponding half-Nyquist filter is given by
\mbox{$G(f)=\sqrt{H(f)}$} where the most prominent example is the
symmetric \ac{RC} and \ac{RRC} pair. The
according non-symmetric CR filter $G\C(f)$ is constructed by
\begin{align}
  G\C(f) &=
  \begin{cases}
    H(f) + j \sqrt{(1-H(f))H(f)} & f\geq 0\\
    H(f) - j \sqrt{(1-H(f))H(f)} & f< 0
  \end{cases}
  \label{eq:CR}
\end{align}
with impulse response $g\C(t)=\mathcal{F}^{-1}\{G\C(f)\}$.  Note that
both $G\C(f)$ and $G\C(f)(G\C(f))^*$ are Nyquist filters
(cf. (\ref{eq:NQ})).

Fig.~\ref{fig:tdfd} shows the \ac{RRC} and \ac{CRRC} filter
response.  The \ac{ICI} $S_1\C(f)$ between adjacent subcarriers with \ac{CR}
filters equals
\begin{align}
  S_1(f) 
  &=\sqrt{H(f)H(f-F)}\\
  S_1\C(f)&=G\C(f-F)[G\C(f)]^*\\ \label{eq:jS1}
  &= jS_1(f),
\end{align}
where (\ref{eq:jS1}) follows from (\ref{eq:CR}) and (\ref{eq:NQ}). Therefore, since
\begin{align}
  s_1\C(t)=js_1(t),
  \label{eq:icitd}
\end{align}
the real and imaginary part of the \ac{ICI} are exchanged when using a \ac{CR}
filter compared to its standard half-Nyquist version.  Both $S\C_1(f)$ and
$s\C_1(t)$ are presented in Fig. \ref{fig:adjcr}.

Hence, if $g(t)$ fulfills (\ref{eq:o1})--(\ref{eq:o4}), the according $g\C(t)$ fulfills
%
\begin{align}
\label{eq:co1} \Re& \bigl\{(g\C(t)w^{-kt}*g\C(-t))|_{t=mT\vphantom{\oh}}\bigr\}&&=\delta(k,m)\\
\label{eq:co2} \Re& \bigl\{(jg\C(t)w^{-kt}*g\C(-t))|_{t=(\oh+m)T}\bigr\}&&=0.
\end{align}
Note, that compared to (\ref{eq:o1})--(\ref{eq:o4}) the factor $j^k$
has been removed.
Consequently, an OFDM/CR-OQAM system that uses a
\ac{CR} filter $g\C(t)$ can be described by the modulation equation
\parbox[t]{\linewidth}{\footnotesize
\begin{align}
  x(t) &= \sum_{k=0\atop m\in\mathbb{Z}}^{K-1}(c_{k,m}^{\text{R}}g\C(t-mT)+jc_{k,m}^{\text{I}}g\C(t-mT-\tfrac{T}{2}))w^{kt},
\end{align}
}
the demodulation equations
\begin{align}
  \hat{c}^{\text{R}}_{k,m} &= \Re (x(t)w^{-kt}*g\C(-t))|_{t=mT}\\
  \hat{c}^{\text{I}}_{k,m} &= \Im (x(t)w^{-kt}*g\C(-t))|_{t=(\oh+m)T}
\end{align}
and the block diagram in Fig. \ref{fig:convOqam} with $\phi=0$.


\begin{figure*}[t]
  \centering
  \subfloat[Time-frequency phase-space.]{\label{fig:phasespace}\includegraphics[height=3.8cm]{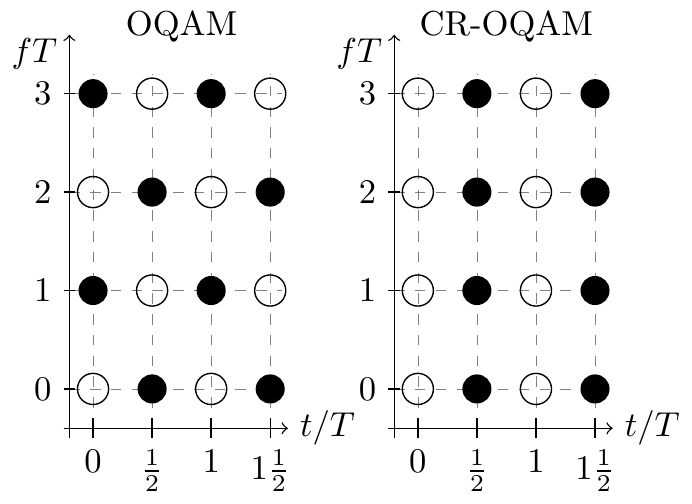}}
  \subfloat[SER performance in fading multipath channels, 16-QAM.]{\label{fig:results}\includegraphics[height=3.8cm]{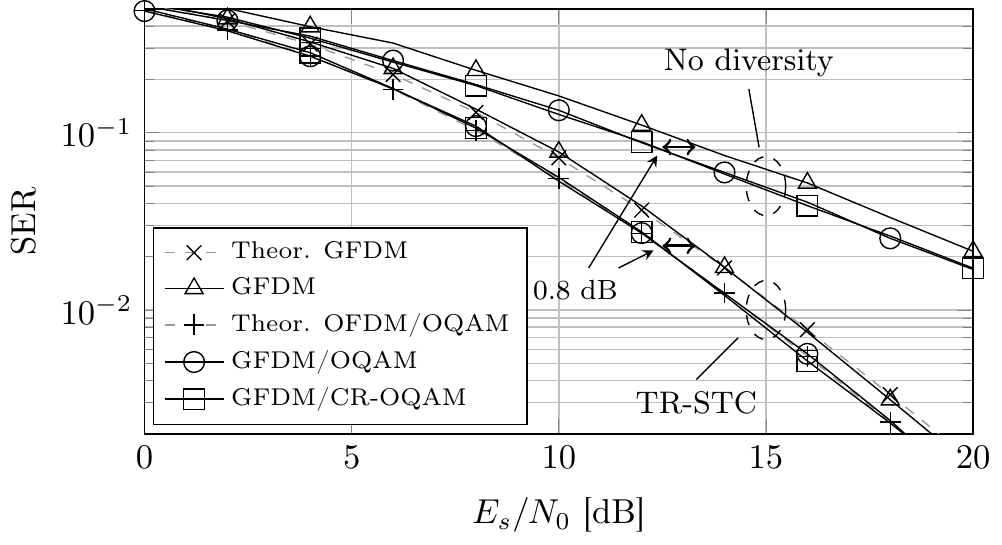}}
  \subfloat[Comparison of PSD.]{\label{fig:spectrum}\includegraphics[height=3.8cm]{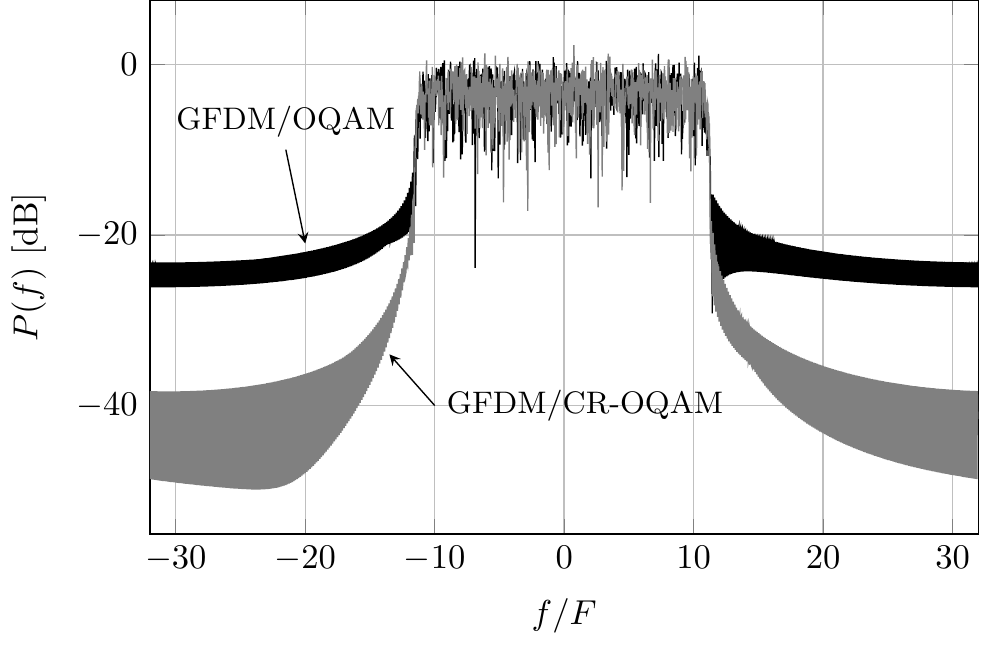}}
   \caption{Time-frequency phase space and performance comparison of OQAM and CR-OQAM.}
\end{figure*}

\tikzexternaldisable The time-frequency phase space for OFDM/OQAM and
OFDM/CR-OQAM is shown in Fig. \ref{fig:phasespace} where real and
imaginary values are depicted with \scalebox{0.7}{\R}~and \scalebox{0.7}{\I}, respectively. Due to
the missing $\tfrac{\pi}{2}$ phase shift between subcarriers, the grid
is more regular for the CR-OQAM case. 
Hence, OFDM/CR-OQAM can be seen as two parallel OFDM/PAM systems
transmitting with a time offset of $\tfrac{T}{2}$ and a
phase shift of $\tfrac{\pi}{2}$.  \tikzexternalenable \MMa[phasespace]{The
  more regular phase space can lead to simpler equalization and pilot
  design and theoretic analysis might be easier to accomplish.}

  \MMa[Laurenti1]{ In \cite{Vangelista2001} conjugate OFDM/OQAM is
    proposed, where forward-time and time-reversed transmit and
    receive filters alternate along the subcarriers, which is clearly
    different compared to the present proposal.  Using this
    approach, the symmetry requirement of the prototype filters is
    relaxed, but still a phase shift between subcarriers is kept,
    hence keeping the phase space equal to conventional OQAM.}\MMa[Laurenti2]{Also in
    \cite{Vangelista2001}, low-complexity formulations for
    conventional and conjugate OQAM are given.  It is stated that due
    to the use of alternating filters for conjugate OQAM, the system
    requires two \acp{PPN} for implementation, which increases the
    complexity.  In the present approach, the symmetry requirement is
    relaxed by using \ac{CR} filters and changing the transmit phase
    space.  Since only one prototype filter type is used, only one
    \ac{PPN} is required, leading to the same implementation
    complexity as conventional OQAM in \cite{Vangelista2001}.  However,
    in \cite{Vangelista2001} the use of one N/2-point complex DFT per
    symbol is required, which would be changed to one N-point real DFT
    per symbol due to the changed phase space for OFDM/CR-OQAM
    implementations.}

\section{Application of CR-OQAM in GFDM}
OFDM/OQAM is a streaming based, filtered \ac{MC} system, where every
symbol overlaps with its adjacent symbols in time.  \ac{GFDM}
\cite{Michailow2014} is a filtered \ac{MC} system, where circular
convolution is applied instead of linear.  Hence, its transmit signal
exhibits a block structure and subsequent blocks can be decoupled by a
\ac{CP} to ease equalization.

\ac{GFDM} is modeled in discrete base band with sampling period
$T_s$. The transmit signal $\vec{x}$ is given by
\begin{align}
  \vec{x}&=\mat{A}\vec{d},
\end{align}
where the columns of the matrix $\mat{A}$ contain circular
time-frequency shifted versions of a prototype transmit filter $g[n]$
with distance $KT_s$ in time and $1/(KT_s)$ in frequency, where $K$ is
the number of subcarriers. $\vec{d}$ contains the complex-valued data
symbols to be transmitted with the block.  By appending a \ac{CP},
frequency domain channel equalization can be carried out at the
receiver, yielding the signal $\hat{\vec{x}}$ and \ac{ZF} or
\ac{MF} detection is applied
\begin{align}
  \hat{\vec{d}}_{\text{ZF}} &= \mat{A}^{-1}\hat{\vec{x}} & \hat{\vec{d}}_{\text{MF}} &= \mat{A}^{\text{H}}\hat{\vec{x}},
\end{align}
where $(\cdot)^\text{H}$ denotes Hermitian conjugate.  A main property
of \ac{GFDM} is its good \ac{TFL} of the transmit filter, which allows to
achieve a low \ac{OOB} radiation and robustness against
asynchronicity \cite{Michailow2014}. However, when using QAM
modulation, the \ac{BLT} prohibits orthogonality completely,
which impacts \ac{MF} performance while \ac{ZF} detectors introduce
noise-enhancement and only exist for certain parameter configurations
\cite{Matthe2014}.  Hence, with perfect synchronization, SER
performance of \ac{GFDM} is worse compared to an orthogonal system.

To circumvent this problem, OQAM modulation can be applied, which
provides orthogonality but still keeps the advantageous
property of good time-frequency localization. Looking at the phase
space of CR-OQAM (Fig. \ref{fig:phasespace}), the GFDM/CR-OQAM
modulator is given by
\begin{align}
  \vec{x}&=\mat{A}\Re\{\vec{d}\}+j \mathcal{C}_{\tfrac{K}{2}}(\mat{A}\Im\{\vec{d}\}),
\end{align}
where $\mathcal{C}_{u}(\cdot)$ denotes a circular rotation of its argument
by $u$ elements. At the receiver, the CR-OQAM detection with the
matched filter is simply achieved by
\begin{align}
  \Re\{\hat{\vec{d}}\}&=\Re\{\mat{A}^H\vec{x}\},&
  \Im\{\hat{\vec{d}}\}&=\Im\{\mat{A}^H\mathcal{C}_{-\tfrac{K}{2}}(\vec{x})\}.
\end{align}
Time-reversal space-time coding (TR-STC) \cite{Al-Dhahir2001} can be
applied to GFDM/CR-OQAM to provide transmit diversity in a fading
multipath environment. TR-STC was initially developed for
single-carrier systems, where two subsequent time domain codewords are
decoupled by a \ac{CP} and jointly space-time encoded.  For \ac{GFDM},
one block is treated as a single-carrier codeword and the
TR-STC is applied onto two subsequent \ac{GFDM} blocks \cite{Matthe2015a}. At the
receiver, both \ac{GFDM} blocks are recovered by
the TR-STC and processed by the \ac{GFDM} detector.  The approach
can be applied to both \ac{GFDM} and GFDM/(CR-)OQAM, but GFDM/(CR-)OQAM is
expected to outperform conventional \ac{GFDM} due to the kept
orthogonality.


\section{Performance evaluation}
The \ac{SER} performance of \ac{GFDM} and GFDM/(CR-)OQAM with and
without TR-STC has been evaluated in Rayleigh fading multipath
channels.  The power delay profile of the channel between the transmit
and receive antennas is modelled with length 16 where the tap powers
decay linearly in log scale from 0dB to -16dB.  The \ac{GFDM}
parameters are given in Tab.  \ref{tab:gfdmconfig} and the simulation
results are presented in Fig. \ref{fig:results}, where perfect channel
state information was available at the receiver.
\MMa[theory]{
The theoretic equations for GFDM/QAM are taken
from \cite{Matthe2015a}.  Note that as GFDM/(CR-)OQAM is orthogonal
it performs equal to OFDM/OQAM and corresponding equations are also
taken from \cite[eq. (19)]{Matthe2015a} with noise-enhancement $\xi=1$.}

GFDM/OQAM outperforms conventional \ac{GFDM} with and without TR-STC due to
the orthogonality introduced by the OQAM modulation. The performance gap equals the
noise-enhancement of \ac{GFDM}, which is $0.8$ dB in the presented
configuration.  Both \ac{GFDM} and GFDM/(CR-)OQAM show a diversity gain of 2
when combined with TR-STC.

\begin{table}[t]
  \centering
  \caption{GFDM simulation parameters}
  \begin{tabular}[t]{llccc}\toprule
   Parameter       & Symbol    & QAM  & OQAM & CR-OQAM \\ \midrule
   Subsymbol spacing  & $K$    & 64   & 64   & 64 \\
   Subsymbol count    & $M$    & 7    &  7   &  7 \\
   Filter          &  $g[n]$   & RC   & RRC  & CRRC \\
   Filter rolloff  & $\alpha$  & 0.5  &  1 & 1 \\
   Detector        &           & ZF   & MF   & MF  \\
   \bottomrule
  \end{tabular}
  \label{tab:gfdmconfig}
\end{table}



\MMa[nonsymmPerf]{Since in the present configuration, OQAM and CR-OQAM
  both use the same equalization scheme, the benefits of
  non-symmetric filters proposed in \cite{Feher1993} cannot be shown
  in the \ac{SER} curve.}
\MMa[spectrum]{ Fig. \ref{fig:spectrum} shows the comparison of the
  \ac{PSD} of GDFM/OQAM and GFDM/CR-OQAM using one \ac{GS}
  \cite{Michailow2014}, where the application of \ac{CRRC}
  filters significantly reduces \ac{OOB} emission. As shown in
  \cite{Michailow2014}, high OOB emission is mainly due to abrupt
  signal changes at the boundaries of the GFDM block. By using
  \acp{GS} with QAM modulation and ISI-free filters,
  these discontinuities are avoided. However, due to the
  half-symbol shift when using OQAM with RRC filters, employing
  \acp{GS} does not remove signal discontinuities. On the other hand,
  as CRRC filters have zeros at twice the symbol rate, application of
  \ac{GS} efficiently reduces OOB emission, making the use of CR-OQAM
  advantageous over that of conventional OQAM.}



\section{Conclusion}
\MMa[conclusion]{
This paper has presented an alternative approach for the
implementation of Offset-QAM when using non-symmetric conjugate root
filters. With the proposal, no phase shift between
subcarriers is required, making the CR-OQAM time-frequency phase
space more regular and compared to existing work, implementation
complexity is reduced.
Orthogonality conditions have been stated for the OFDM/CR-OQAM system
and it was proven that \ac{CR} filters fulfill these.
CR-OQAM has been applied to GFDM to create an orthogonal
system with good \ac{TFL}. Space-time code aiming for transmit diversity
was applied to GFDM/(CR-)OQAM and, outperforming the
conventional space-time coded GFDM system.  \ac{GS} were
efficiently employed to reduce \ac{OOB} emission when using CR
filters, which is not possible with conventional OQAM.}


\bibliographystyle{IEEEtran}
\bibliography{library}

\end{document}